\documentclass[twocolumn]{article}
\newtheorem{theorem}{Theorem}

\usepackage{latexsym} 

\input{epsf}

\usepackage{amsmath}
\usepackage{amssymb}
\usepackage{amsfonts}
\usepackage{graphicx}

\def\smallromani{\renewcommand{\theenumi}{\roman{enumi}}
\renewcommand{\labelenumi}{(\theenumi)}}

\newcommand{\oldbfe}[1]{\begin{bfseries}\emph{#1}\end{bfseries}}

\newcommand{\ES}{\mbox{$\emptyset$}}

\newcommand{\myra}{\mbox{$\:\rightarrow\:$}}

\newcommand{\A}{\mbox{$\ \wedge\ $}}

\newcommand{\LL}{\mbox{$\ldots$}}

\newcommand{\C}[1]{\mbox{$\{{#1}\}$}}           

\newcommand{\NI}{\noindent}

\newcommand{\VV}{\vspace{5 mm}}

\newcommand{\szkew}[1]{\relax \setbox0=\hbox{\kern -24pt $\displaystyle#1$\kern 0pt }%
\box0}
{\catcode`\@=11 \global\let\ifjusthvtest@=\iffalse}

\newcounter{oldmycaption}

\usepackage{color}
\usepackage{sgame}
\usepackage{amsmath}
\usepackage{colordvi}
\usepackage{latexsym}

\input xy 
\xyoption{all} 

\def\defemb#1#2{\expandafter\def\csname #1\endcsname 
  {\relax\ifmmode #2\else\hbox{$#2$}\fi}}

\defemb{cA}{{\cal A}} 
\defemb{cB}{{\cal B}} 
\defemb{cC}{{\cal C}} 
\defemb{cD}{{\cal D}} 
\defemb{cE}{{\cal E}} 
\defemb{cG}{{\cal G}} 
\defemb{cI}{{\cal I}} 
\defemb{cK}{{\cal K}} 
\defemb{cN}{{\cal N}} 
\defemb{cM}{{\cal M}} 
\defemb{cP}{{\cal P}} 
\defemb{cR}{{\cal R}} 
\defemb{cS}{{\cal S}} 
\defemb{cT}{{\cal T}} 
\defemb{cV}{{\cal V}} 
\defemb{cO}{{\cal O}} 

\defemb{bF}{{\bf F}} 
\defemb{bp}{{\bf p}} 
\defemb{bU}{{\bf U}} 
\defemb{bu}{{\bf u}} 
\defemb{bV}{{\bf V}} 
\defemb{bW}{{\bf W}} 
\defemb{bw}{{\bf w}} 
\defemb{bX}{{\bf X}} 
\defemb{bx}{{\bf x}} 
\defemb{bY}{{\bf Y}} 
\defemb{by}{{\bf y}} 
\defemb{bZ}{{\bf Z}} 
\defemb{bz}{{\bf z}}

\newcommand{\mycomment}[1]{}

\def\0{{\bf 0}} 

\def\1{{\bf 1}}

\newcommand{\commentout}[1]{}

\newcommand{\comment}[1]{}

\def\0{{\bf 0}}
\def\1{{\bf 1}}

 \makeatletter
\def\@normalsize{\@setsize\normalsize{12pt}\xpt\@xpt
\abovedisplayskip 11pt plus2pt minus5pt\belowdisplayskip
\abovedisplayskip \abovedisplayshortskip \z@
plus3pt\belowdisplayshortskip 6pt plus3pt
minus3pt\let\@listi\@listI}

\def\subsize{\@setsize\subsize{12pt}\xipt\@xipt}
\def\section{\@startsection{section}{1}{\z@}{24pt plus 2 pt
minus 2 pt} {12pt plus 2pt minus 2pt}{\large\bf}}
\def\subsection{\@startsection {subsection}{2}{\z@}{12pt
plus 2pt minus 2pt}{12pt plus 2pt minus 2pt}{\subsize\bf}}
\makeatother

\begin{document}
\date{}
\title{\Large\bf CP-nets and Nash equilibria}
\author{Krzysztof R. Apt$^{1,2,3}$, Francesca Rossi$^{4}$, and Kristen Brent Venable$^{4}$ \\
$^{1}$ School of Computing, National University of Singapore \\
$^{2}$ CWI, Amsterdam \\ 
$^{3}$University of Amsterdam, the Netherlands \\
$^{4}$Department of Pure and Applied Mathematics, University of Padova
}
\maketitle

\thispagestyle{empty}
\subsection*{\centering Abstract}
\vspace*{-3mm} {\it
  
  We relate here two formalisms that are used for different
  purposes in reasoning about multi-agent systems. One of them are
  strategic games that are used to capture the idea that agents
  interact with each other while pursuing their own interest.  The
  other are CP-nets that were introduced to express 
  qualitative and conditional preferences of the users and which aim
  at facilitating the process of preference elicitation.
  
  To relate these two formalisms we introduce a natural, qualitative,
  extension of the notion of a strategic game.  We show then that the
  optimal outcomes of a CP-net are exactly the Nash equilibria of an
  appropriately defined strategic game in the above sense. This allows
  us to use the techniques of game theory to search for optimal
  outcomes of CP-nets and vice-versa, to use techniques developed for
  CP-nets to search for Nash equilibria of the considered games.  }

\section{Introduction}


One of the main tools in the area of multi-agent systems is game
theory, notably strategic games. They formalize in a simple and
powerful way the idea that agents interact with each other while
pursuing their own interest.  The interaction is captured by the fact
that actions (strategies) are taken simultaneously, while agents'
interests are expressed by means of the utility (payoff) function that
each agent wishes to maximize.

\emph{CP-nets} (Conditional Preference nets) are an elegant formalism
for representing conditional and qualitative preferences, see
\cite{BBHP.UAI99,BBHP.journal}.  They model such
preferences under a {\em ceteris paribus} (that is, `all else being
equal') assumption.  The CP-net represents a complex
`joint preference distribution' in a compact form.
Preference elicitation in such a framework
appears to be natural and intuitive.  

Research on CP-nets focused on its modeling capabilities and
algorithms for solving various natural problems related to their use.
Also, computational complexity of these problems was extensively
studied.  An outcome of a CP-net is an assignment of values to its
variables. One of the fundamental problems is that of finding an
optimal outcome, i.e., the one that cannot be improved in presence of
the adopted preference statements.  This is in general a complex
problem since it was found that finding optimal outcomes and testing
for their existence is NP-hard in general. In contrast, for so-called
acyclic CP-nets this is an easy problem which can be solved by a
linear time algorithm.

The aim of this paper to show the relationship between CP-nets and
game theory, and explain how game-theoretic techniques developed for the
analysis of strategic games can be fruitfully used to study CP-nets.
To this end, we introduce a generalization of the customary strategic
games (see, e.g., \cite{Mye91},) in which each player has to
his disposal a strict preference relation on his set of strategies,
parametrized by a joint strategy of his opponents. We call such games
\emph{strategic games with parametrized preferences}.

The cornerstone of our approach are two results closely relating
CP-nets to such games. They show that the optimal outcomes of a CP-net
are exactly the Nash equilibria of an appropriately defined strategic
game with parametrized preferences. This allows us to transfer
techniques of game theory to CP-nets. 



To find Nash equilibria in strategic games, reduction techniques
have been studied which reduce the game by eliminating some players'
strategies, thus obtaining a smaller game. We introduce two
counterparts of such game-theoretic techniques that allow us to reduce
a CP-net while maintaining its optimal outcomes. We also introduce a
method of simplifying a CP-net by eliminating so-called redundant
variables from the variables parent sets. Both techniques simplify the
search for optimal outcomes of a CP-net.


The paper is organized as follows.  Section \ref{back} provides the
basic definitions of CP-nets.  Then, Section \ref{sec:games}
introduces our generalized notion of games, Section \ref{sec:to-games}
shows how to pass from a CP-net to a game, and Section
\ref{sec:to-nets} handles the opposite direction.  Then, Section
\ref{sec:reduced} introduces the concept of reduced CP-nets, and
Section \ref{sec:games-to-nets} shows how to exploit techniques
developed in games for CP-nets.  Finally, Section \ref{conc}
summarizes the main contributions of the paper and discusses current and
future work.

\section{CP-nets}
\label{back}


CP-nets~\cite{BBHP.UAI99,BBHP.journal} (for Conditional Preference nets)
are a graphical model for
compactly representing conditional and qualitative preference
relations. They exploit conditional preferential independence by
decomposing an agent's preferences via the {\bf ceteris paribus} (cp)
assumption. Informally, CP-nets are sets of \oldbfe{ceteris 
paribus (cp)\/} preference statements. For instance, the statement
{\em ``I prefer red wine to white wine if meat is served."} asserts
that, given two meals that differ {\em only} in the kind of wine
served {\em and} both containing meat, the meal with a red wine is
preferable to the meal with a white wine.
On the other hand, this statement does not order 
two meals with a different main course. 
Many users' preferences 
appear to be of this type.

CP-nets bear some similarity to Bayesian networks. Both utilize
directed graphs where each node stands for a domain variable, and
assume a set of \emph{features} (variables) $F = \{X_1,\ldots,X_n\}$
with the corresponding finite domains $\cal D$$(X_1),\ldots,$$\cal
D$$( X_n)$.  For each feature $X_i$, a user specifies a (possibly
empty) set of \oldbfe{parent} features $Pa(X_i)$ that can affect her
preferences over the values of $X_i$.  This defines a
\oldbfe{dependency graph} in which each node $X_i$ has $Pa(X_i)$ as
its immediate predecessors.  

Given this structural information, the
user explicitly specifies her preference over the values of $X_i$ for
{\em each complete assignment} on $Pa(X_i)$. This preference is
assumed to take the form of a linear ordering over $\cal D$$(X_i)$
\cite{BBHP.UAI99,BBHP.journal}.  Each such specification is called
below a \oldbfe{preference statement} for the variable $X_i$.  
These conditional preferences over the values of $X_i$ are
captured by a {\em conditional preference table} which is
annotated with the node $X_i$ in the CP-net.
An \oldbfe{outcome} is an assignment of values to the variables with each
value taken from the corresponding domain.

As an example, consider a CP-net 
whose features are $A$, $B$, $C$ and $D$, with binary domains
containing $f$ and $\overline{f}$ if $F$ is the name of the feature,
and with the following preference statements: 

$d: a \succ \overline{a}$, \ $\overline{d} : a \succ \overline{a}$,

$a : b \succ \overline{b}$, \ $\overline{a} : \overline{b} \succ b$, 

$b : c \succ \overline{c}$, \ $\overline{b} : \overline{c} \succ c$, 

$c : d \succ \overline{d}$, \ $\overline{c} : \overline{d} \succ d$.

\NI
Here the preference statement $d: a \succ \overline{a}$
states that $A=a$ is preferred to $A=\overline{a}$, given that $D = d$.
From the structure of these preference statements we see that
$Pa(A) = \C{D}, Pa(B) = \C{A}, Pa(C) = \C{B}, Pa(D) = \C{C}$
so the dependency graph is cyclic.

An \oldbfe{acyclic} CP-net is one in which the dependency graph is
acyclic. As an example, consider a CP-net
whose features and domains are as above
and with the following preference statements: 

$a \succ \overline{a}$, 

$b \succ \overline{b}$, 

$(a \wedge b) \vee (\overline{a} \wedge \overline{b}) : c \succ \overline{c}$, \ 
$(a \wedge \overline{b}) \vee (\overline{a} \wedge b) : \overline{c} \succ c$, 

$c: d \succ \overline{d}$, \ $\overline{c}: \overline{d} \succ d$.  

\NI
Here, the preference statement $a \succ \overline{a}$ represents the
unconditional preference for $A=a$ over $A=\overline{a}$.  
Also each preference statement for the variable $C$ 
is a actually an abbreviated version of two preference statements.
In this example we
have $Pa(A) = \ES, Pa(B) = \ES, Pa(C) = \C{A,B}, Pa(D) = \C{C}$.

The semantics of CP-nets depends on the notion 
of a \oldbfe{worsening flip}. A
worsening flip is a transition between two outcomes that consists
of a change in the value of a single variable to one which
is less preferred in the unique preference statement for that
variable.
By analogy we define an \oldbfe{improving flip}.
For example, in the acyclic CP-net above,
passing from $abcd$ to $ab\overline{c}d$ is a worsening flip since $c$
is better than $\overline{c}$ given $a$ and $b$.  We say that an
outcome $\alpha$ is \oldbfe{better} than the outcome $\beta$ 
(or, equivalently, $\beta$ is \oldbfe{worse} than $\alpha$),  
written as $\alpha \succ \beta$, iff there is a chain 
of worsening flips from $\alpha$ to $\beta$.
This definition induces a strict preorder over the outcomes.
In the above acyclic CP-net 
the outcome 
$\overline{a}b\overline{c}\overline{d}$ is worse than $abcd$.

An \oldbfe{optimal} outcome is 
one for which no better outcome exists.
In general, a CP-net does not need to have an optimal outcome. As an
example consider two features $A$ and $B$ with the respective domains
$\C{a, \overline{a}}$ and $\C{b, \overline{b}}$ and the following
preference statements:

$a: b \succ \overline{b}$, \ $\overline{a}: \overline{b} \succ b$,

$b: \overline{a} \succ a$, \ $\overline{b}: a \succ \overline{a}$.

\NI
It is easy to see that then 

$
ab \succ a\overline{b} \succ \overline{a}\overline{b} \succ \overline{a}b \succ ab.
$

\section{Strategic games with parametrized preferences}
\label{sec:games}

In this section we introduce a generalization of 
the notion of a strategic game
used in game theory, see, e.g., \cite{Mye91}.

First we need the concept of a \oldbfe{preference} on a set $A$ which
in this paper denotes a strict linear ordering on $A$. If $\succ$ is a
preference, we denote by $\succeq$ the corresponding \oldbfe{weak
preference} defined by: $a \succeq b$ iff $a \succ b$ or $a = b$.

Given a sequence of non-empty sets $S_1, \LL, S_n$ and
$s \in S_1 \times \LL \times S_n$ we denote the $i$th element of $s$ by $s_i$ and
use the following standard notation of game theory, where $I := i_1, \LL, i_k$
is a subsequence of $1, \LL, n$:

\begin{itemize}
\item $s_{-i} := (s_1, \LL, s_{i-1}, s_{i+1}, \LL, s_n)$,

\item $s_{I} := (s_{i_1}, \LL, s_{i_k})$,

\item $(s'_i, s_{-i}) := (s_1, \LL, s_{i-1}, s'_i, s_{i+1}, \LL, s_n)$, where
we assume that $s'_i \in S_i$,

\item $S_{-i} := S_1 \times \LL \times S_{i-1} \times S_{i+1} \times \LL \times S_n$,

\item $S_{I} := S_{i_1} \times  \LL \times  S_{i_k}$.
\end{itemize}

In game theory it is customary to study strategic games in which the
outcomes are numerical values provided by means of the payoff
functions.  A notable exception is \cite{OR94} in which instead of
payoff functions the linear quasi-orderings on the sets of joint
strategies are used.

In our setup we adopt a different approach according to which each
player has to his disposal a strict preference relation $\succ
\hspace{-1.5mm}(s_{-i})$ on his set of strategies \emph{parametrized}
by a joint strategy $s_{-i}$ of his opponents.  So in our approach
\begin{itemize}

\item  for each $i \in [1..n]$ player $i$ has
a finite, non-empty, set $S_i$ of strategies available to him,

\item for each $i \in [1..n]$ and $s_{-i} \in S_{-i}$
player $i$ has a preference relation $\succ\hspace{-1.5mm}(s_{-i})$
on his set of strategies $S_i$. 
\end{itemize}

In what follows such a \oldbfe{strategic game with parametrized
  preferences} (in short a \oldbfe{game with parametrized
  preferences}, or just a \oldbfe{game}) for $n$ players is
represented by a sequence
\[
(S_1, \LL, S_n, {\succ}(s_{-1}), \LL, {\succ}(s_{-n})),
\] 
where each $s_{-i}$ ranges over $S_{-i}$.

It is straightforward to transfer to the case of games with
parametrized preferences the basic notions concerning strategic games.
In particular, given a game
$G$ with parametrized preferences specified as above we say that
a joint strategy $s$ is a (pure) \oldbfe{Nash equilibrium} of $G$ if 
for all $i \in [1..n]$ and all $s'_i \in S_i$
\[
s_i \succeq\hspace{-1.5mm}(s_{-i}) \ s'_i.
\]
(For the original definition see, e.g., \cite{Mye91}.)

To clarify this definitions consider the classical
Prisoner's dilemma strategic game represented by the following
bimatrix representing the payoffs to both players:
\begin{center}
\begin{game}{2}{2}
       & $C_2$    & $N_2$\\
$C_1$   &$3,3$   &$0,4$\\
$N_1$   &$4,0$   &$1,1$
\end{game}
\end{center}
So each player $i$ has two strategies, $C_i$ (cooperate) and $N_i$ (not cooperate),
the payoff to player 1 for the joint strategy $(C_1,N_2)$ is 0, etc.
To represent this game as a game with parametrized preferences we simply
stipulate that

$\succ \hspace{-1.5mm}(C_2) := \ N_1 \succ C_1$, \ 
$\succ \hspace{-1.5mm}(N_2) := \ N_1 \succ C_1$,

$\succ \hspace{-1.5mm}(C_1) := \ N_2 \succ C_2$, \ 
$\succ \hspace{-1.5mm}(N_1) := \ N_2 \succ C_2$.

\NI
These orderings reflect the fact that for each strategy of the
opponent each player considers his `not cooperate' strategy better
than his `cooperate' strategy.
It is easy to check that
$(N_1, N_2)$ is a unique Nash equilibrium of this game with 
parametrized preferences.

\section{From CP-nets to strategic games}
\label{sec:to-games}

Consider now a CP-net with the set of variables $\C{X_1, \LL, X_n}$
with the corresponding finite domains ${\cal D}(X_1), \LL, {\cal D}(X_n)$.
We write each preference statement for the variable $X_i$ as 
$X_I = a_I : \ \succ_i$,
where for the subsequence $I = i_1, \LL, i_k$ of $1, \LL, n$:
\begin{itemize}
\item $Pa(X_i) = \C{X_{i_1}, \LL, X_{i_k}}$,
\item $X_I = a_I$ is an abbreviation for
$X_{i_1} = a_{i_1} \A \LL \A X_{i_k} = a_{i_k}$,
\item $\succ_i$ is a preference over ${\cal D}(X_i)$.
\end{itemize}
We also abbreviate ${\cal D}(X_{i_1}) \times \LL \times {\cal D}(X_{i_k})$
to ${\cal D}(X_I)$.

By definition, the preference statements for a variable $X_i$
are exactly all statements of the form 
$X_I = a_I: \ \succ\hspace{-1.5mm}(a_I)$, 
where $a_I$ ranges over ${\cal D}(X_I)$ and 
$\succ\hspace{-1.5mm}(a_I)$ is a preference on 
${\cal D}(X_i)$ that depends on $a_I$.

We now associate with each 
CP-net $N$ a game ${\cal G}(N)$ with parametrized preferences
as follows:
\begin{itemize}
\item each variable $X_i$ corresponds to a player $i$,
\item the strategies of player $i$ are the elements of the domain ${\cal D}(X_i)$
of $X_i$.
\end{itemize}

To define the parametrized preferences, consider a player $i$. Suppose
$Pa(X_i) = \C{X_{i_1}, \LL, X_{i_k}}$ and let $I := i_1, \LL, i_k$. So
$I$ is a subsequence of $1, \LL, i-1, i+1, \LL, n$.  Given a joint
strategy $a_{-i}$ of the opponents of player $i$, we associate with it
the preference relation 
$\succ\hspace{-1.5mm}(a_{I})$ on ${\cal D}(X_i)$ where
$X_I = a_I: \  \succ\hspace{-1.5mm}(a_{I})$ is the 
unique preference statement
for $X_i$ determined by $a_I$.

In words, the preference of a player $i$ over his strategies, 
given the strategies chosen by its opponents, say $a_{-i}$,  
coincides with the preference given by the CP-net over the domain of 
$X_i$ given the assignment to his parents $a_{I}$ which must coincide
with the projection of $a_{-i}$ over $I$.
This completes the definition of ${\cal G}(N)$.

As an example  consider the first CP-net of Section \ref{back}.
The corresponding game has four players $A$, $B$, $C$, $D$, each with
two strategies indicated with $f$, $\bar{f}$ for player $F$.
The preference of each player on his strategies will depend only on the
strategies chosen by the players which correspond to his parents in the
CP-net. 
Consider for example player $B$. His preference
over his strategies $b$ and $\bar{b}$,
given the joint strategy of his opponents $s_{-B}=dac$, is 
$b \succ \bar{b}$. Notice that, for example, the
same ordering holds for the opponents joint strategy
$s_{-B}=\bar{d}a\bar{c}$, since the strategy chosen by the only player
corresponding to his parent, $A$, has not changed.     
 
We have then the following result.

\begin{theorem} \label{thm:G(N)}
An outcome of a CP-net $N$ is optimal iff it is 
a Nash equilibrium of the game ${\cal G}(N)$.
\end{theorem}

\section{From strategic games to CP-nets}
\label{sec:to-nets}

We now associate with each game $G$ with parametrized preferences 
a CP-net ${\cal N}(G)$ as follows:

\begin{itemize}
\item each player $i$ corresponds to a variable $X_i$,  
\item the domain ${\cal D}(X_i)$ of the variable $X_i$ consists of
  the set of strategies of player $i$,
\item we stipulate that $Pa(X_i) = \C{X_{1}, X_{i-1}, \LL, X_{i+1}, \LL, X_n}$,
where $n$ is the number of players in $G$.
\end{itemize}

Next, for each joint strategy $s_{-i}$ of 
the opponents of player $i$ we take
the preference statement
$X_{-i} = s_{-i}: \  \succ\hspace{-1.5mm}(s_{-i})$,
where $\succ\hspace{-1.5mm}(s_{-i})$ is the preference relation on the set of strategies
of player $i$ associated with $s_{-i}$.

This completes the definition of ${\cal N}(G)$.  As an example of this
construction let us return to the Prisoner's dilemma game with
parametrized preferences from Section \ref{sec:games}.  
In the corresponding CP-net we have then
two variables $X_1$ and $X_2$ corresponding to players 1 and 2, with
the respective domains $\C{C_1, N_1}$ and $\C{C_2, N_2}$.  To explain how
each parametrized preference translates to a preference statement take
for example $\succ \hspace{-1.5mm}(C_2) := \ N_1 \succ C_1$. It
translates to $X_2 = C_2 : \ N_1 \succ C_1$.

\NI
We have now the following counterpart of Theorem \ref{thm:G(N)}.

\begin{theorem} \label{thm:N(G)}
A joint strategy is a Nash equilibrium of the game $G$ iff it is an optimal
outcome of the CP-net ${\cal N}(G)$.
\end{theorem}

\section{Reduced CP-nets}
\label{sec:reduced}

The disadvantage of the above construction of the CP-net ${\cal N}(G)$ from a
game $G$ is that it always produces a CP-net in which all sets of
parent features are of size $n-1$ where $n$ is the number of features
of the CP-net.  This can be rectified by reducing each set of parent
features to a minimal one as follows.

Given a CP-net $N$, consider a variable $X_i$ with the parents 
$Pa(X_i)$, and take a variable $Y \in Pa(X_i)$. 
Suppose that for all assignments $a$ to
$Pa(X) - \{Y\}$ and any two values $y_1,y_2 \in {\cal D}(Y)$, 
the orderings  $\succ\hspace{-1.5mm}(a,y_1)$ and $\succ\hspace{-1.5mm}(a,y_2)$
on ${\cal D}(X_i)$ coincide.

We say then that $Y$ is \emph{redundant} in the set of parents of $X_i$.
It is easy to see that by removing all redundant variables 
from the set of parents of $X_i$ and 
by modifying the corresponding preference statements for $X_i$ accordingly, 
the strict preorder $\succ$ over the outcomes of the CP-nets is not changed.

Given a CP-net, if for all its variable $X_i$ the set  $Pa(X_i)$ 
does not contain any redundant variable, we say that the CP-net is 
\oldbfe{reduced}.

By iterating the above construction every CP-net can be transformed to
a reduced CP-net. 
As an example consider a CP-net with three features,
$X, Y$ and $Z$, with the respective domains $\C{a_1, a_2}, \C{b_1,
 b_2}$ and $\C{c_1, c_2}$.  Suppose now that $Pa(X) = Pa(Y) = \ES,
Pa(Z) = \C{X, Y}$ and that

$\succ\hspace{-1.5mm}(a_1, b_1) = \ \succ\hspace{-1.5mm}(a_2, b_1)$, \ 
$\succ\hspace{-1.5mm}(a_1, b_2) = \ \succ\hspace{-1.5mm}(a_2, b_2)$, 

$\succ\hspace{-1.5mm}(a_1, b_1) = \ \succ\hspace{-1.5mm}(a_1, b_2)$, \ 
$\succ\hspace{-1.5mm}(a_2, b_1) = \ \succ\hspace{-1.5mm}(a_2, b_2)$.

\NI
Then both $X$ and $Y$ are redundant, so we can reduce the CP-net
by 
reducing $Pa(Z)$ to $\emptyset$.
$Z$ becomes an independent variable in the reduced CP-net with an
ordering over 
its domain which coincides with the unique one given in the original CP-net in
terms of the assignments to its parents.


%
%




In what follows for a CP-net $N$ we denote by 
$r(N)$ the corresponding 
reduced CP-net.
The following result summarizes the relevant properties of $r(N)$ and
relates it to the constructions of ${\cal G}(N)$ and ${\cal N}(G)$.

\begin{theorem} \label{thm:reduced}
\mbox{} \vspace{-3mm}

\begin{enumerate} \smallromani

\item Each CP-net $N$ and its reduced form 
$N' = r(N)$
have the same ordering $\succ$ over the outcomes. 

\item For each CP-net $N$ and its reduced form $N' = r(N)$ we have
${\cal G}(N) = {\cal G}(N')$. 

\item Each reduced CP-net $N$ is a 
reduced CP-net corresponding to the game ${\cal G}(N)$.
Formally: $N = r({\cal N}({\cal G}(N)))$.
\end{enumerate} 
\end{theorem}
\VV

Part $(i)$ states that the reduction procedure 
preserves the ordering over the outcomes.
Part $(ii)$ states that the construction of a game corresponding to 
a CP-net does not depend on the redundancy of the given CP-net.
Finally, part $(iii)$
states that the reduced CP-net $N$ can be obtained `back' from the game 
${\cal G}(N)$.

Games $G$ such that the CP-net $N' = r({\cal N}(G))$ is acyclic
are not uncommon. In fact, they naturally represent multi-agent scenarios
where
agents (that is, players of the game)
can be partitioned into levels $1,2,\ldots, n$, such that 
agents at level $i$ can express their
preferences (that is, payoff function) without looking at what players at
higher levels do.
Informally, agents at level $i$ are more important than
agents at level $j$ is $j>i$.
In particular, agents at level $1$ can decide their
preferences without looking at the behavior of any other agent.

\section{Game-theoretic techniques in CP-nets}
\label{sec:games-to-nets}

Given the correspondence between CP-nets and games
and its properties presented in the previous sections, we 
can now use them to transfer standard techniques
of game theory, used to find Nash equilibria, to CP-nets to find 
their optimal outcomes.

More specifically, we can transfer two techniques 
of iterated elimination of `suboptimal' strategies ---those that
are strictly dominated or are never best responses
(see, e.g., \cite{OR94}.)
To introduce them in the context of CP-nets
consider a CP-net $N$
with the set of variables $\C{X_1, \LL, X_n}$
with the corresponding finite domains ${\cal D}(X_1), \LL, {\cal D}(X_n)$.

\begin{itemize}

\item We say that an element $d_i$ from 
the domain $\cal D$$(X_i)$ of the variable $X_i$
is a \oldbfe{best response} to a preference statement
\[
X_I = a_I : \ \succ_i
\]
for $X_i$ 
if $d_i \succeq_i d'_i$ 
for all $d'_i \in {\cal D}(X_i)$.

\item We say that an element $d_i$ from the domain of the variable $X_i$
is a \oldbfe{never a best response}
if it is not a best response to any preference statement for $X_i$.

\item Given two elements $d_i, d'_i$ from the domain $\cal D$$(X_i)$
  of the variable $X_i$ we say that $d'_i$ is \oldbfe{strictly
    dominated} by $d_i$ if for all preference statements $X_I = a_I :
  \ \succ_i$ for $X_i$ we have
\[
d_i \succ_i d'_i.
\] 

\end{itemize}

By a \oldbfe{subnet} of a CP-net $N$ we mean a CP-net obtained from $N$
by removing some elements from some variable domains followed by the removal
of all preference statements that refer to a removed element.

Then we introduce the following 
relation between a CP-net $N$ and its subnet $N'$:
\[
N \myra_{\hspace{-1mm} NBR \: } N'
\]
when $N \neq N'$ and for each variable $X_i$ 
each removed element from the domain of $X_i$ is never a best response
in $N$, and introduce an analogous relation $N \myra_{\hspace{-1mm} S \: } N'$
for the case of strictly dominated elements.

The following result then holds.
\begin{theorem}
\label{thm:nets-ienbr}

Suppose that $N \myra^{*}_{\hspace{-1mm} NBR} N'$, i.e., the CP-net $N'$ is
obtained by an iterated elimination of never best responses from the
CP-net $N$.  

  \begin{enumerate} \smallromani
  \item Then $s$ is an optimal outcome of $N$ iff it is an optimal outcome of $N'$.
    
  \item If each variable in $N'$ has a singleton domain, then the resulting outcome is a 
    unique optimal outcome of $N$.  

  \item All iterated eliminations of never best responses from the
CP-net $N$ yield the same final outcome. 

  \end{enumerate}
\end{theorem}

To illustrate the use of this theorem reconsider the first CP-net from
Section \ref{back}, i.e., the one with the preference statements

$d: a \succ \overline{a}$, \ $\overline{d} : a \succ \overline{a}$,

$a : b \succ \overline{b}$, \ $\overline{a} : \overline{b} \succ b$, 

$b : c \succ \overline{c}$, \ $\overline{b} : \overline{c} \succ c$, 

$c : d \succ \overline{d}$, \ $\overline{c} : \overline{d} \succ d$.

\NI
Denote it by $N$.

We can reason about it using the iterated elimination of strictly dominated strategies
(which coincides here with the iterated elimination  of never best responses, since each
domain has exactly two elements).

We have the following chain of reductions:
\[
N \myra_{\hspace{-1mm} S} N_1 \myra_{\hspace{-1mm} S} N_2 \myra_{\hspace{-1mm} S} N_3 \myra_{\hspace{-1mm} S} N_4,
\]
where

\begin{itemize}
\item $N_1$ results from $N$ by removing $\overline{a}$ (from the domain of $A$) and the 
preference statements $d: a \succ \overline{a}$, \ $\overline{d} : a \succ \overline{a}$, \ 
$\overline{a} : \overline{b} \succ b$, 

\item $N_2$ results from $N_1$ by removing $\overline{b}$ and the 
preference statements
$a : b \succ \overline{b}$, \  $\overline{b} : \overline{c} \succ c$, 

\item $N_3$ results from $N_2$ by removing $\overline{c}$ and the 
preference statements $b : c \succ \overline{c}$ \ $\overline{c} : \overline{d} \succ d$, 

\item $N_4$ results from $N_3$ by removing $\overline{d}$ from the domain of $D$ and the 
preference statement $c : d \succ \overline{d}$.

\end{itemize}

Indeed, in each step the removed element is strictly dominated in the
considered CP-net.  So using the iterated elimination of strictly
dominated elements we reduced the original CP-net to one in which each
variable has a singleton domain and consequently found a unique
optimal outcome of the original CP-net $N$.

Finally, the following result shows that the introduced reduction relation
on CP-nets is complete for acyclic CP-nets.

\begin{theorem}
For each acyclic CP-net $N$ a unique subnet $N'$ with the singleton domains exists such that
$N \myra^{*}_{\hspace{-1mm} NBR} N'$.
\end{theorem}

\section{Conclusions and future work}
\label{conc}

In this paper we related two formalisms that are commonly used in
reasoning about multi-agent systems, strategic games and CP-nets.  To
this end we generalized the concept of strategic games to games with
parametrized preferences and showed that optimal outcomes in CP-nets
are exactly Nash equilibria of such games.  This allowed us to exploit
game-theoretic techniques in search for optimal outcomes of CP-nets.

Our current research deals two other aspects concerning strategic
games and preferences.  First, thanks to the established
correspondence, we can also use the techniques developed to reason
about optimal outcomes of a CP-net in search for Nash equilibria of
strategic games with parametrized preferences. These techniques, as
recently shown in \cite{dimo,aaai05}, involve the use of the customary
constraint solving techniques.  In fact, it has been shown that the
optimal outcomes of any CP-net, even a cyclic one, can be found by
just solving a set of hard constraints. Thus hard constraint solving
is enough to find also Nash equilibria in strategic games.

Second, we found that the direct correpondence between the optimal
solutions of a CP-net and the Nash equilibria of the corresponding
game cannot be easily found in other preference modelling formalisms,
for example soft constraints, see \cite{jacm}.  In fact, it is
possible to show that, in a so-called fuzzy constraint problem, there
can be optimal solutions which are not Nash equilibria of
corresponding games, and vice-versa.  We are therefore studying the
conditions under which soft constraints can be
related to game theory.

In this paper we assumed that payoff functions give a linear order
over the strategies of a player. It could be useful to see whether our
results can be generalized to games in which players' strategies can
be incomparable or indifferent to each other, thus using partial
orderings with ties.  We are currently studying this scenario.

This paper is just a first step towards what we think is a 
fruitful cross-fertilization between preferences, constraint solving, 
and game theory. 



\bibliographystyle{plain}

\bibliography{nash}

\end{document}